\documentclass{aa}  

\usepackage{graphicx, amsmath, array}
\usepackage{color}
\usepackage[normalem]{ulem}
\usepackage{hyperref}

\usepackage{xcolor}

\newcommand{\refs}[1]{\textcolor{red}{REFS}}

\newcommand{\Msun}{M_{\odot}}
\newcommand{\mpeak}{\textsc{Multi Peak}\xspace}
\newcommand{\cpeak}{\textsc{Compactness Peaks}\xspace}

\usepackage{natbib}
\bibpunct{(}{)}{;}{a}{}{,} 

\begin{document}

\title{Compactness peaks: An astrophysical interpretation of the mass distribution of merging binary black holes}

\titlerunning{Astrophysical interpretation of the BBH mass distribution}


\author{Shanika Galaudage
          \inst{1,2}\thanks{\email{shanika.galaudage@oca.eu}}
          \and
          Astrid Lamberts\inst{1,2}
          }
\institute{Laboratoire Lagrange, Université Côte d'Azur, Observatoire de la Côte d'Azur, CNRS,  Bd de l'Observatoire, 06300, France
    \and
        Laboratoire Artemis, Université Côte d'Azur, Observatoire de la Côte d'Azur, CNRS,  Bd de l'Observatoire, 06300, France\\
    }

\date{Received MONTH DD, YYYY; accepted MONTH DD, YYYY}

\abstract{With the growing number of detections of binary black hole mergers, we are beginning to probe structure in the distribution of masses.
A recent study by Schneider et al. proposes that isolated binary evolution of stripped stars naturally gives rise to the peaks at $\mathcal{M} \sim 8 \Msun$, $14 \Msun$ in the chirp mass distribution and explains the dearth of black holes between $\mathcal{M}\approx 10-12 \Msun$.
The gap in chirp mass results from an apparent gap in the component mass distribution between $m_1, m_2 \approx 10-15 \Msun$ and the specific pairing of these black holes.
This component mass gap results from the variation in core compactness of the progenitor, where a drop in compactness of Carbon-Oxygen core mass will no longer form black holes from core collapse.
We develop a population model motivated by this scenario to probe the structure of the component mass distribution of binary black holes consisting of two populations: 1) two peak components to represent black holes formed in the \textit{compactness peaks}, and 2) a powerlaw component to account for any \textit{polluting events}, these are binaries that may have formed from different channels (e.g. dynamical).
We perform hierarchical Bayesian inference to analyse the events from the third gravitational-wave transient catalogue (GWTC-3) with this model.
We find that there is a preference for the lower mass peak to drop off sharply at $\sim 11 \Msun$ and the upper mass peak to turn on at $\sim 13 \Msun$, in line with predictions from Schneider et al. 
However, there is no \textit{clear} evidence for a gap.
We also find mild support for the two populations to have different spin distributions.
In addition to these population results, we highlight observed events of interest that differ from the expected population distribution of compact objects formed from stripped stars.} 


\maketitle

\section{Introduction}

Gravitational-wave astronomy is a rapidly growing field with $\sim 90$ detections \citep{GWTC3:KAGRA:2021vkt} reported by the LIGO-Virgo-KAGRA collaboration (LVK). 
This catalogue of events was produced with data from the Advanced LIGO \citep{LIGO:LIGOScientific:2014pky}, Advanced Virgo \citep{VIRGO:2014yos} gravitational-wave observatories.
The majority of these detections are of merging binary black holes.
With population studies of these sources, the structure of the mass distribution of binary black holes is slowly being revealed. This structure has implications for stellar physics processes such as the explosion mechanisms of supernovae and mass transfer processes in binaries \citep[see][for reviews]{Mapelli:2018uds, Mandel:2018hfr, Mandel:2021smh, Spera:2022byb}.

Majority of the focus thus far in terms of astrophysical interpretation of these merging binaries has been in relation to the components masses of the compact binaries, with statistically significant peaks in the distribution at $\approx 10 \Msun$ and $\approx 35 \Msun$ in the primary mass distribution \citep[e.g.][]{Talbot:2018cva, Sadiq:2021fin, Edelman:2021zkw, Wong:2022bxp, Farah:2023vsc, GWTC3popKAGRA:2021duu}.
The former peak may be associated with binary mass transfer processes \citep[e.g.][]{vanSon:2022myr}, while the latter could be related to stellar physics such as pulsational pair instability supernova \citep[PPISN;][]{Heger:2001cd, Croon:2023kct, Rahman:2021dvs}, but there are suggestions that the fraction of events in the peak is too high to be from PPISN \citep[e.g.][]{Stevenson:2019rcw}, or the location of peak due to PPISN should be higher \citep[e.g.][]{Farmer:2020xne, Farag:2022jcc, Hendriks:2023yrw, Golomb:2023vxm}, or the peak is populated by merging binaries in globular clusters \citep{Antonini:2022vib}.
However, recent studies have started to look beyond the component mass distributions to investigate and interpret the chirp mass distribution in more detail \citep[e.g.][]{GWTC3popKAGRA:2021duu,Tiwari:2023xff, Schneider:2023mxe}.

The chirp mass \citep{Finn:1992xs, Poisson:1995ef, Blanchet:1995ez} determines the leading-order orbital evolution of a binary system from gravitational-wave emission and thus is measured most accurately,
\begin{align}
    \mathcal{M} = \frac{(m_1 m_2)^{3/5}}{(m_1 + m_2)^{1/5}}. 
\end{align}
It is a combination of the individual component masses $m_1$ and $m_2$ and is directly related to gravitational-wave frequency and frequency evolution.
The chirp-mass distribution of binary black hole mergers shows statistically significant peaks at 8, 14 and 28 $\Msun$ with an apparent lack of binary black hole mergers with $\mathcal{M} \approx 10 - 12 \Msun$ \citep[e.g.][]{GWTC3popKAGRA:2021duu,Tiwari:2023xff}, henceforth referred to as the ``chirp mass gap''.

A recent study, \cite{Schneider:2023mxe} proposed that merging binary black holes formed via isolated binary evolution where 
the progenitors lost their hydrogen-rich envelopes through mass transfer processes naturally give rise to the peaks at $\mathcal{M} \sim 8 \Msun$, $14 \Msun$ and a dearth of black holes between $\mathcal{M}\approx 10-12 \Msun$.
They find that the bimodality in binary black hole chirp-mass distribution is due to the pairing of ``compactness peaks'' in the component mass distribution which is linked to stellar physics and supernova mechanisms.
 
The explodability of stars is dependant on the compactness of the stellar core, $\xi = M/R(M)$, where $M$ is the mass of the core, and $R(M)$ is the radius at mass coordinate $M$, where $M=2.5 \Msun$ is used in 
\cite{Schneider:2023mxe}.
The compactness is generally considered to \textit{not} steadily increase with core mass, but rather it has sharp peaks and dips due to variations in burning stages in the core \citep[e.g.][]{OConnor:2010moj, Sukhbold:2013yca, Patton:2020tiy, Chieffi:2020gxh, Schneider:2020vvh, Schneider:2023mxe}, however, there are studies in which a monotonic increase in compactness is considered \citep{Mapelli:2019ipt}.
Considering the binary stripped stars in the \cite{Schneider:2023mxe} models, for 10\% solar metallicity, the compactness level peaks due to neutrino-dominated burning of Carbon-Oxygen (CO) at core masses ($M_\mathrm{CO}$) of $7.5 \Msun$, producing black holes from core collapse supernova at $\approx 9-10 \Msun$; these are the low mass black holes, which we will refer to as $\mathrm{BH_L}$, the same terminology as used in \cite{Schneider:2023mxe}.
Above this point, the compactness in the core drops, falling into a region forming neutron stars from core collapse instead of black holes (see Figure 3 in \citealt{Schneider:2023mxe}).
At around $M_\mathrm{CO} \approx 13 \Msun$ burning becomes neutrino dominated in the Oxygen-Neon (ONe) stage as well, producing black holes from core collapse supernova greater than $15 \Msun$; these are the high mass black holes, which we will refer to as $\mathrm{BH_H}$.
This variation in compactness level results in no black hole masses being formed between $\approx 10-15 \Msun$.
These boundaries can shift due to uncertainties in the stellar physics, including metallicity dependant mass-loss rate \citep[][]{Vink:2005zf, Belczynski:2010ApJ...714.1217B, Spera:2015MNRAS.451.4086S};  convective core boundary mixing during core hydrogen and core helium burning \citep[e.g.][]{Schneider:2023mxe, Temaj:2023nuo} and $^{12}$C($\alpha, \gamma$)$^{16}$O nuclear reaction rates \cite[e.g.][]{Farmer:2019ApJ...887...53F, Farmer:2020xne, Costa:2021MNRAS.501.4514C}.\footnote{See Figure 6 in \cite{Schneider:2023mxe} for an illustration of how physical processes shift the edges of the $10-15 \Msun$ gap in component mass.}
However, the shifts at the lower mass end are expected to be within $\approx 1 \Msun$.

The peaks in chirp mass are produced by a mixture of pairings between $\mathrm{BH_L}$ and $\mathrm{BH_H}$.
The peak below the chirp mass gap is formed by $\mathrm{BH_L} + \mathrm{BH_L}$ and the structure above the chirp mass gap are formed by $\mathrm{BH_H} + \mathrm{BH_H}$. Any $\mathrm{BH_L} + \mathrm{BH_H}$ binaries are expected to fill this chirp mass gap, but the pairing of $\mathrm{BH_L} + \mathrm{BH_H}$ is less likely due to isolated binary evolution preferring to produce closer to equal mass ratio systems \citep{Schneider:2020vvh}.
Assuming this model hypothesis in correct, the $\mathrm{BH_L} + \mathrm{BH_H}$ population may be completely suppressed according to current observations \citep{GWTC3:KAGRA:2021vkt}.

In this work, we develop a parametric population model motivated by the Schneider et al. results to make an astrophysical interpretation of mass distribution. In Section \ref{sec:methods} we provide a description of the population model and analysis method. In Section \ref{sec:results} we present our findings using our model in comparison to commonly used models by the LVK. Finally in Section \ref{sec:discussion} we discuss the astrophysical implications of our results and avenues for future work.

\section{Methodology}\label{sec:methods}

\begin{figure*}
    \centering
    \includegraphics[width=\textwidth]{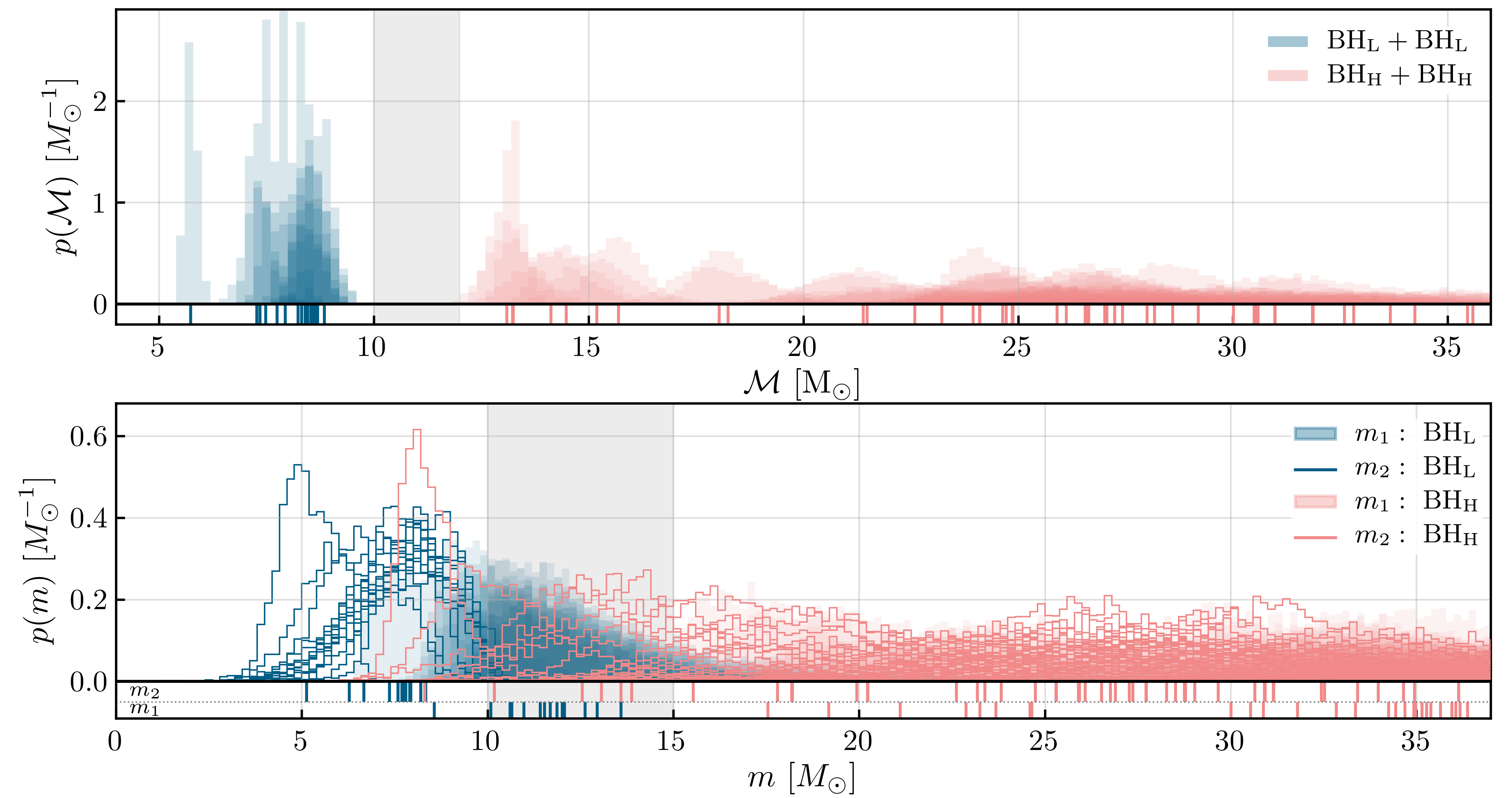}
    \caption{Events in GWTC-3 with FAR $< 1 \mathrm{yr^{-1}}$ sorted by which chirp mass peak they lie in assuming the \cite{Schneider:2023mxe} predictions on their classification and location of the gaps. \textit{Top panel}: Chirp mass posteriors showing the $\mathrm{BH_L} + \mathrm{BH_L}$ (blue) and the $\mathrm{BH_H} + \mathrm{BH_H}$ (pink). The grey shaded region represents the chirp mass gap region ($\mathcal{M} \approx 10 - 12 \Msun$). \textit{Bottom panel}: Component mass posteriors showing the $\mathrm{BH_L}$ (blue) and the $\mathrm{BH_H}$ (pink) black holes. The grey shaded region represents the expected component mass gap region ($m \approx 10 - 15 \Msun$). The vertical ticks illustrate the median of each posterior.}
    \label{fig:mchirpm1m2_sorted}
\end{figure*}

\begin{figure}
    \centering
    \includegraphics[width=0.99\columnwidth]{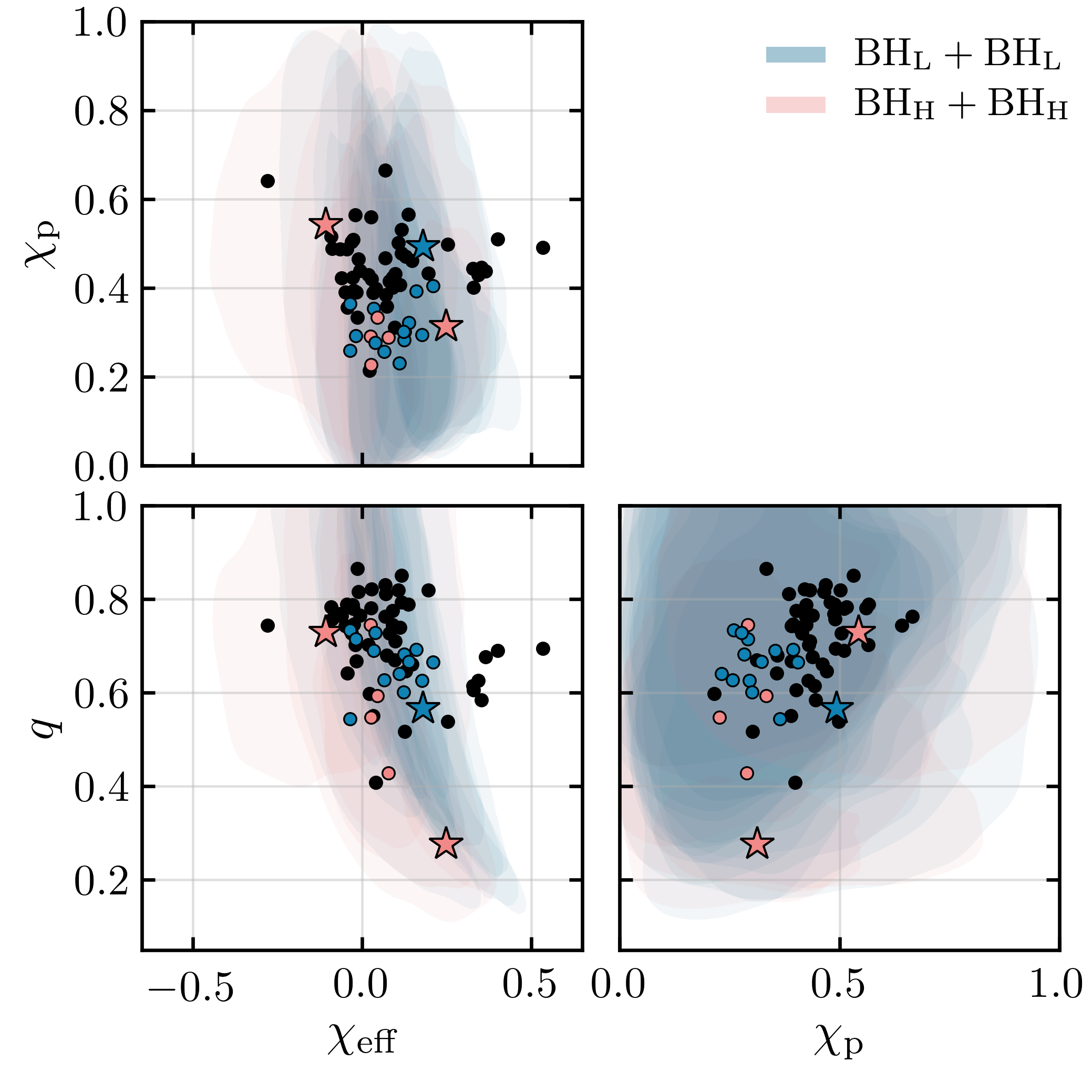}
    \caption{Two-dimensional distributions of mass ratio ($q$), effective inspiral spin ($\chi_\mathrm{eff}$) and effective precession spin ($\chi_p$) of events polluting the $m=10-15 \Msun$ gap (pink and blue shaded region; 90\% credibility) and the median (pink and blue dots). The median of the posteriors of the events outside the gap are also shown (black dots).
    Events polluting the gap (GW190412, GW151226 and GW200225), that may have formed via a channel different to isolated binary evolution, are highlighted with stars.}
    \label{fig:m1m2_filtered}
\end{figure}

In this work, we explore whether the findings in \cite{Schneider:2023mxe} have any explanatory power for the population of observed binary black hole mergers.
To do this, we first investigate the properties of the events individually in Section \ref{subsec:methods_individual}.
Then we develop an astrophysically-informed population model in Section \ref{subsec:methods_population} to investigate the properties of the population and whether there is a gap-like feature in the component mass distribution.

\subsection{Identifying binary black hole systems of interest}\label{subsec:methods_individual}

We consider the binary black hole mergers analysed in previous population studies \citep[e.g.][]{GWTC3popKAGRA:2021duu} which include 69 binary black hole mergers from the third LVK gravitational-wave transient catalogue GWTC-3 \citep{GWTC3:KAGRA:2021vkt} detected with a FAR $< 1 \mathrm{yr^{-1}}$.

As an initial step, we investigate which part of the chirp mass distribution the events lie in and sort them by whether they reside in the low or high mass region as illustrated in the upper panel of Figure 
\ref{fig:mchirpm1m2_sorted}.
We assume that any events in the low mass region ($\mathcal{M} < 11 \Msun$) are  $\mathrm{BH_L} + \mathrm{BH_L}$ mergers and any in the high mass region ($\mathcal{M} > 11 \Msun$)  are $\mathrm{BH_H} + \mathrm{BH_H}$ mergers.
We then show where these objects lie in the component mass space.
We find that from this first inspection, while there is a clear gap in the chirp mass distribution, the component mass distribution does not show a gap-like feature.

Of course, a visual inspection of the posteriors does not necessarily mean there is no gap-like structure, especially given the uncertainties on the mass posteriors being wider than the width of the gap.
A population study, as outlined in Section \ref{subsec:methods_population}, is required to investigate this structure.
However, it is interesting to consider whether there are binaries ``polluting'' the $\approx 10-15 \Msun$ region in component mass that were formed via other channels.

We identify the events where the median of the mass posterior distributions (for one or more components) lies above $10 \Msun$ for events with $\mathcal{M} < 11 \Msun$ and below $10 \Msun$ for events with $\mathcal{M} > 11 \Msun$.
We find 20 events that pollute the component mass gap and list some of the properties in Table \ref{tab:eoi}, including mass ratio ($q=m_2/m_1$), effective inspiral spin ($\chi_\mathrm{eff}$) and effective precession spin ($\chi_p$).\footnote{The effective inspiral spin parameter, $\chi_\mathrm{eff}$, is a mass-weighted spin parameter that is projected along the orbital angular momentum, and the effective precession spin parameter, $\chi_p$, is a measure of in-plane spin, which enables the parameterisation of the rate of relativistic precession of the orbital plane}.
These parameters may suggest the binary formed via a different channel.
Unequal mass ratio mergers are unlikely to occur in isolated binary evolution \citep[e.g.][]{Kruckow:2018slo, Giacobbo:2018etu, Broekgaarden:2021efa, Iorio:2022sgz,  Schneider:2020vvh}.
Negative values $\chi_\mathrm{eff}$ and large values of $\chi_p$ are generally considered to be a signature of dynamical formation given the support for misaligned and in-plane spin orientations \citep{Mandel:2009nx, Rodriguez:2016vmx, Zevin:2017evb, Rodriguez:2017pec, Doctor:2019ruh}.

We also indicate whether the events have been considered in other studies to have formed via a channel that is different to isolated binary evolution.
To obtain a visual sense of these potential ``pollutants'' (i.e. events that may not have formed via isolated binary evolution) we also provide the $q$, $\chi_\mathrm{eff}$ and $\chi_p$ distributions for these events in Figure \ref{fig:m1m2_filtered} along with the median of posterior for the other 49 events, illustrated by the black dots.

There are a few events in the component mass gap (Table \ref{tab:eoi}) that appear to differ from the typical properties of systems formed via isolated evolution with more extreme mass ratios and higher values of $\chi_p$ (e.g. GW190412, GW151226, GW200225) compared to the rest of the population.
However, there are also events not in this mass gap that have these differing properties too, (Figure \ref{fig:m1m2_filtered}).
Given this information, we need to build a population model that can account for potential pollutants in the population in addition to finding the location and edges of the compactness peaks.
Finding this structure will help determine whether or not there is a gap-like feature between $\approx 10-15 \Msun$ in the component mass distribution of binary black hole mergers.

\begin{table*}
\centering
 \renewcommand{\arraystretch}{1.5}
 \begin{tabular}{c | c | c  | c  | p{8cm}  } 
    \hline
    {\bf Event}; [REF] & q & $\chi_\mathrm{eff}$ & $\chi_\mathrm{p}$  & {\centering Suggestions for other formations scenarios}
    \\\hline\hline 
GW190412\_053044; [1]& $0.28^{+0.12}_{-0.06}$ & $0.25^{+0.08}_{-0.11}$ & $0.31^{+0.2}_{-0.16}$ & \footnotesize{e.g. \citet{Safarzadeh:2020qrc, Gerosa:2020bjb, Liu:2020gif}} \\
GW151226; [2] & $0.57^{+0.37}_{-0.34}$ & $0.18^{+0.2}_{-0.12}$ & $0.5^{+0.39}_{-0.33}$ & \footnotesize{e.g. \citet{Chatterjee:2016thb}}\\
GW190828\_065509; [3] & $0.43^{+0.38}_{-0.16}$ & $0.08^{+0.15}_{-0.16}$ & $0.29^{+0.41}_{-0.21}$ & - \\
GW200316\_215756; [4] & $0.6^{+0.34}_{-0.39}$ & $0.12^{+0.28}_{-0.1}$ & $0.29^{+0.4}_{-0.21}$ & - \\
GW190728\_064510; [3] & $0.68^{+0.27}_{-0.31}$ & $0.12^{+0.13}_{-0.06}$ & $0.28^{+0.32}_{-0.2}$ & - \\
GW190720\_000836; [3] & $0.63^{+0.31}_{-0.27}$ & $0.17^{+0.13}_{-0.11}$ & $0.3^{+0.35}_{-0.21}$ &  - \\
GW191204\_171526; [4] & $0.7^{+0.26}_{-0.26}$ & $0.16^{+0.08}_{-0.05}$ & $0.39^{+0.36}_{-0.26}$ & - \\
GW190708\_232457; [3] & $0.75^{+0.22}_{-0.28}$ & $0.02^{+0.1}_{-0.08}$ & $0.3^{+0.43}_{-0.24}$ &  -\\
GW190930\_133541; [3] & $0.66^{+0.28}_{-0.37}$ & $0.14^{+0.2}_{-0.14}$ & $0.33^{+0.38}_{-0.23}$ &  -\\
GW191216\_213338; [4] & $0.63^{+0.31}_{-0.28}$ & $0.11^{+0.13}_{-0.06}$ & $0.23^{+0.33}_{-0.16}$ & - \\
GW190707\_093326; [3] & $0.73^{+0.22}_{-0.22}$ & $-0.03^{+0.08}_{-0.07}$ & $0.25^{+0.38}_{-0.2}$ & - \\
GW190512\_180714; [3] & $0.54^{+0.37}_{-0.18}$ & $0.03^{+0.13}_{-0.13}$ & $0.23^{+0.35}_{-0.18}$ & - \\
GW191103\_012549; [4] & $0.67^{+0.29}_{-0.36}$ & $0.21^{+0.16}_{-0.1}$ & $0.4^{+0.4}_{-0.28}$ & - \\
GW170608; [5] & $0.7^{+0.27}_{-0.37}$ & $0.03^{+0.19}_{-0.07}$ & $0.35^{+0.44}_{-0.27}$ & - \\
GW190725\_174728; [3] & $0.57^{+0.38}_{-0.31}$ & $-0.04^{+0.26}_{-0.14}$ & $0.37^{+0.46}_{-0.29}$ & - \\
GW200225\_060421; [4] & $0.73^{+0.24}_{-0.28}$ & $-0.11^{+0.16}_{-0.28}$ & $0.53^{+0.34}_{-0.38}$ & \footnotesize{e.g. \citet{Antonelli:2023gpu, Zhang:2023fpp}} \\
GW191105\_143521; [4] & $0.71^{+0.25}_{-0.3}$ & $-0.02^{+0.13}_{-0.09}$ & $0.29^{+0.46}_{-0.24}$ & - \\
GW151012; [6] & $0.59^{+0.36}_{-0.35}$ & $0.05^{+0.3}_{-0.21}$ & $0.34^{+0.45}_{-0.26}$ & - \\
GW191129\_134029; [4] & $0.63^{+0.31}_{-0.3}$ & $0.06^{+0.16}_{-0.08}$ & $0.26^{+0.37}_{-0.2}$ & - \\
GW200202\_154313; [4] & $0.72^{+0.24}_{-0.31}$ & $0.04^{+0.12}_{-0.06}$ & $0.28^{+0.41}_{-0.21}$ & -\\ 
    \hline
  \end{tabular}
  \caption{
  Properties of events ``polluting'' the component mass gap ($m = 10-15 \Msun$). We provide the 90\% credible intervals for the mass ratio ($q$), effective inspiral spin ($\chi_\mathrm{eff}$) and effective precession spin ($\chi_p$) of events for which the median of the mass posterior distributions (for one or more components) lies above $10 \Msun$ for events with $\mathcal{M} \gtrapprox 11 \Msun$ and below $10 \Msun$ for events with $\mathcal{M} \gtrapprox 11 \Msun$. The events are ordered from having the most to least posterior support in the gap. Included are the publication where the event is first reported: [1] \citet{GW190412:LIGOScientific:2020stg}; [2] \citet{GW151226:LIGOScientific:2016sjg}; [3] \citet{GWTC2:LIGOScientific:2020ibl}; [4] \citet{GWTC3:KAGRA:2021vkt}; [5] \citet{GW170608:LIGOScientific:2017vox}; [6] \citet{GWTC0:LIGOScientific:2016dsl}; and some studies that have suggested a formation scenario different to isolated binary evolution for a given system. 
  }
  \label{tab:eoi}
\end{table*}

\subsection{Building an astrophysically-motivated population model}\label{subsec:methods_population}

We develop an astrophysically-informed population model incorporating the features predicted in \citeauthor{Schneider:2023mxe} for the component mass distributions while accounting for ``pollutant'' events.
We perform hierarchical Bayesian inference in order to measure the population hyper-parameters using gravitational-wave data from the LVK.\footnote{See \cite{Thrane:2018qnx} for a detailed review on parameter estimation and hierarchical inference in gravitational wave astronomy.}
We use \texttt{GWPopulation} \citep{GWPopulation:Talbot:2019okv}, a hierarchical Bayesian inference package which employs the use of \texttt{Bilby} \citep{Bilby:Ashton:2018jfp, BilbyGWTC1:Romero-Shaw:2020owr}. We use the nested sampler \texttt{DYNESTY} \citep{dynesty:Speagle:2019ivv} for our analyses.
Our dataset contains the 69 BBH observations from GWTC-3 \citep{GWTC3:KAGRA:2021vkt} that were considered reliable for population analyses \citep[events with a false alarm rate $< 1 \mathrm{yr}^{-1}$;][]{GWTC3popKAGRA:2021duu}. 
We use the posterior samples from parameter estimation results used in \cite{GWTC3popKAGRA:2021duu} along with the injection sets to account for selection effects \citep{Essick:2022ojx}.
There are three components to our model with various hyperparameters ($\Lambda$):

\begin{itemize}
    \item A low mass narrow Gaussian ($\mathrm{BH_L}$ Peak): this models the ``compactness peak'' where black holes are formed from increased compactness of CO cores of stars from neutrino-dominated burning,
    \begin{align}
        \pi(m_1 |\Lambda)_\mathrm{BH_L} = \mathcal{N}(m_1| \mu_\mathrm{BH_L},\sigma_\mathrm{BH_L},m_\mathrm{min}^{\mathrm{BH_L}},m_\mathrm{max}^{\mathrm{BH_L}}),
    \end{align}
    where $\mu_\mathrm{BH_L}$ and $\sigma_\mathrm{BH_L}$ are the mean and width for the Gaussian and $m_\mathrm{min}^{\mathrm{BH_L}}$ and $m_\mathrm{max}^{\mathrm{BH_L}}$ are the lower and upper limits of the Gaussian. 
    \item A high mass broad Gaussian ($\mathrm{BH_H}$ Peak): this models the black holes formed the increased compactness from neutrino release in the ONe burning stage,
    \begin{align}
        \pi(m_1 |\Lambda)_\mathrm{BH_H} = \mathcal{N}(m_1| \mu_\mathrm{BH_H},\sigma_\mathrm{BH_H},m_\mathrm{min}^{\mathrm{BH_H}},m_\mathrm{max}^{\mathrm{BH_H}}),
    \end{align}
    where $\mu_\mathrm{BH_H}$ and $\sigma_\mathrm{BH_H}$ are the mean and width for the Gaussian and $m_\mathrm{min}^{\mathrm{BH_H}}$ and $m_\mathrm{max}^{\mathrm{BH_H}}$ are the lower and upper limits of Gaussian. 
    \item A power law component (\textsc{PL}): this models everything that does not fit within these peaks, possibly from black holes formed via other channels,
    \begin{align}
        \pi(m_1 |\Lambda)_\mathrm{PL} \propto 
        \begin{cases}
        m_1^{-\alpha} & m_\mathrm{min}^\mathrm{PL} < m_1 < m_\mathrm{max}^\mathrm{PL} \\
        0 & \text{otherwise}
        \end{cases}
    \end{align}
    where $\alpha$ is the slope index of the powerlaw and and $m_\mathrm{min}^{\mathrm{BL}}$ and $m_\mathrm{max}^{\mathrm{PL}}$ are the lower and upper truncation points of the powerlaw.
\end{itemize}
Each of these components is paired with different mass ratio distributions, where,
\begin{align}
    \pi(q | m_1, \Lambda)_X \propto 
    \begin{cases}
    q^{{\beta}_\mathrm{X}} & m_\mathrm{min}^\mathrm{X} < m_2 < m_1 \\
    0 & \text{otherwise},
    \end{cases}
    \label{eq:mass_ratio}
\end{align}
where $X = \mathrm{BH_L}$ is the mass ratio distribution for the low mass Gaussian component, $X = \mathrm{BH_H}$ is for the high mass Gaussian component, and $X = \mathrm{PL}$ is for the power law component.
For each component, the mass ratio is constrained such that $m_2$ also has to be larger than the minimum mass of that given component $m_\mathrm{min}$.

We consider the fraction of BBH in each component, where $f_\mathrm{peaks}$ is the fraction of BBH in the $\mathrm{BH_L}$ and $\mathrm{BH_H}$ peaks, and $f_1$ is the fraction of BBH in $\mathrm{BH_L}$ peak compared to the total number of BBH in the two peaks.
Therefore the fraction of BBH in the $\mathrm{BH_L}$ peak, $\mathrm{BH_H}$ peak and powerlaw components are $f_{\mathrm{BH_L}} = f_\mathrm{peaks} \times f_1$, $f_{\mathrm{BH_H}} = f_\mathrm{peaks} \times (1-f_1)$ and $f_{\mathrm{PL}} = 1 - f_\mathrm{peaks}$ respectively.
Details on the prior ranges of these parameters are provided in \ref{tab:compactnesspeaks_priors}.

Considering our model, we may naively expect that the peaked distributions may have different spin properties to the powerlaw component.
The peaked distributions are from black holes forming via isolated evolution, hence this population would generally be expected to form BBH systems with spins preferentially aligned with the orbital angular momentum.
Supernova kicks may misalign the spins, but the misalignment angle is generally considered to be modest~\citep{OShaughnessy:2017eks,Stevenson:2017dlk,Gerosa:2018wbw,Rodriguez:2016vmx,Bavera:2020inc}.
For black holes in the power law component, they may have formed through different channels, some of which may have formed dynamically. 
These systems are expected to form black hole binaries with isotropically distributed spin orientations ~\citep{Mandel:2009nx, Rodriguez:2016vmx, Zevin:2017evb, Rodriguez:2017pec, Doctor:2019ruh}.
If we have hierarchical mergers, we may also expect to see greater spin magnitudes in this population \citep[e.g.][]{Yang:2019cbr, Gerosa:2017kvu}.

We consider two spin populations, one for the BBH in the peaks ($X = \mathrm{peaks}$), and one for the BBH in the powerlaw $X = \mathrm{PL}$. The spin magnitude distribution is modelled by a truncated Gaussian, 

\begin{align}\label{eq:spin_mag_model}
    \pi(\chi_{1,2} | \Lambda)_X = \prod_{i=1}^{2}\mathcal{N}(\chi_i|\mu_\chi^X,\sigma_\chi^X) ,
\end{align}

where $\mu_\chi$ and $\sigma_\chi$ are the mean and width of the Gaussian, and the distribution is truncated at $\chi=0$ and $\chi=1$. The spin orientation model is modelled by a mixture model of a uniform distribution and truncated Gaussian \citep{Talbot:2017yur},
\begin{align}\label{eq:spin_tilt_model}
    \pi( \cos{\theta}_{1,2} | \Lambda)_X =~ & \frac{1-\zeta^X}{4} + \zeta^X \prod_{i=1}^{2}\mathcal{N}(\cos\theta_i|\sigma_t^X),
\end{align}
where $\theta$ is the angle of misalignment of a component with respect to the orbital angular momentum, $\sigma_t$ is the width of the Gaussian, and the mean is fixed to 1. The distribution is truncated at $\cos\theta=-1$ and $\cos\theta=1$. 

Henceforth we will refer to the model with these mass and spin prescriptions as \cpeak.
In addition to masses and spins, we also fit for the redshift distribution using the \textsc{PowerLaw} redshift evolution \citep{Fishbach:2018edt} used in \cite{GWTC3popKAGRA:2021duu} for easy comparison with results from LVK analyses. 
We assume the same redshift distribution for each component of the model ($\pi( z | \Lambda)$), and hence do not discuss these analysis results.\footnote{Supplementary material including analysis inputs, posterior samples and additional plots are available here \href{https://github.com/shanikagalaudage/bbh_compactness_peaks}{https://github.com/shanikagalaudage/bbh\_compactness\_peaks}}

\section{Results}\label{sec:results}

\begin{figure}
    \centering
    \includegraphics{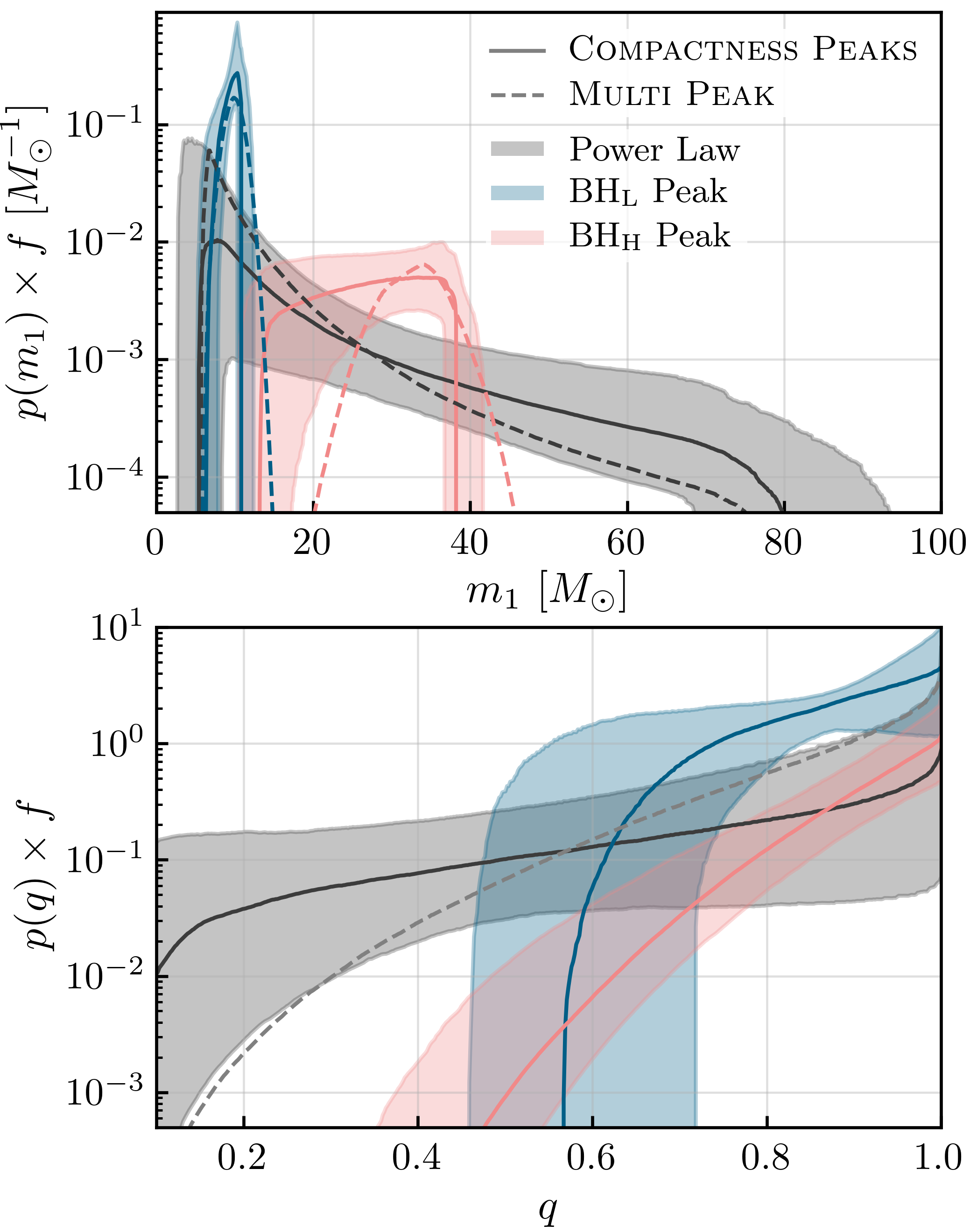}
    \caption{Population distributions for primary mass ($m_1$, top panel) and mass ratio ($q$, bottom panel) using the \cpeak model for binary black holes mergers scaled by the fraction of events in the  $\mathrm{BH_L}$ Peak (blue), $\mathrm{BH_H}$ Peak (pink) and \textsc{Power Law} (black) components.
    The solid curve is the median of the distribution and the shaded regions represent the 90\% credible interval.
    The dashed line represents the median of the distribution for the \mpeak model, which has the same mass ratio distribution for all components.}
    \label{fig:mass_dist}
\end{figure}

\begin{figure}
    \centering
    \includegraphics{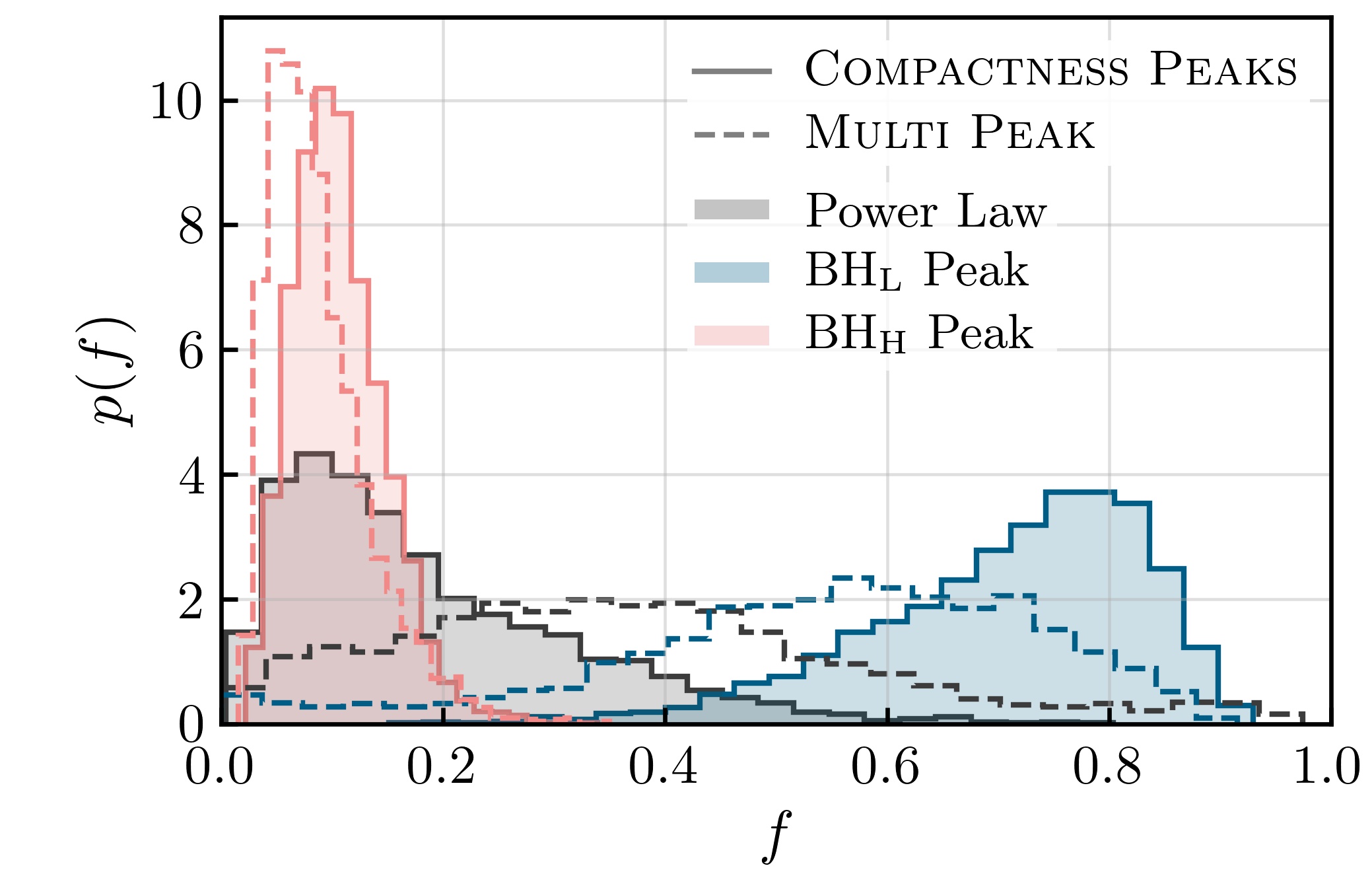}
    \caption{Fraction of binary black holes in the  $\mathrm{BH_L}$ Peak (blue), $\mathrm{BH_H}$ Peak (pink) and \textsc{Power Law} (black) components using the \cpeak (solid lines) and \mpeak (dashed lines) model analyses. }
    \label{fig:fracs}
\end{figure}

\begin{figure}
    \centering
    \includegraphics{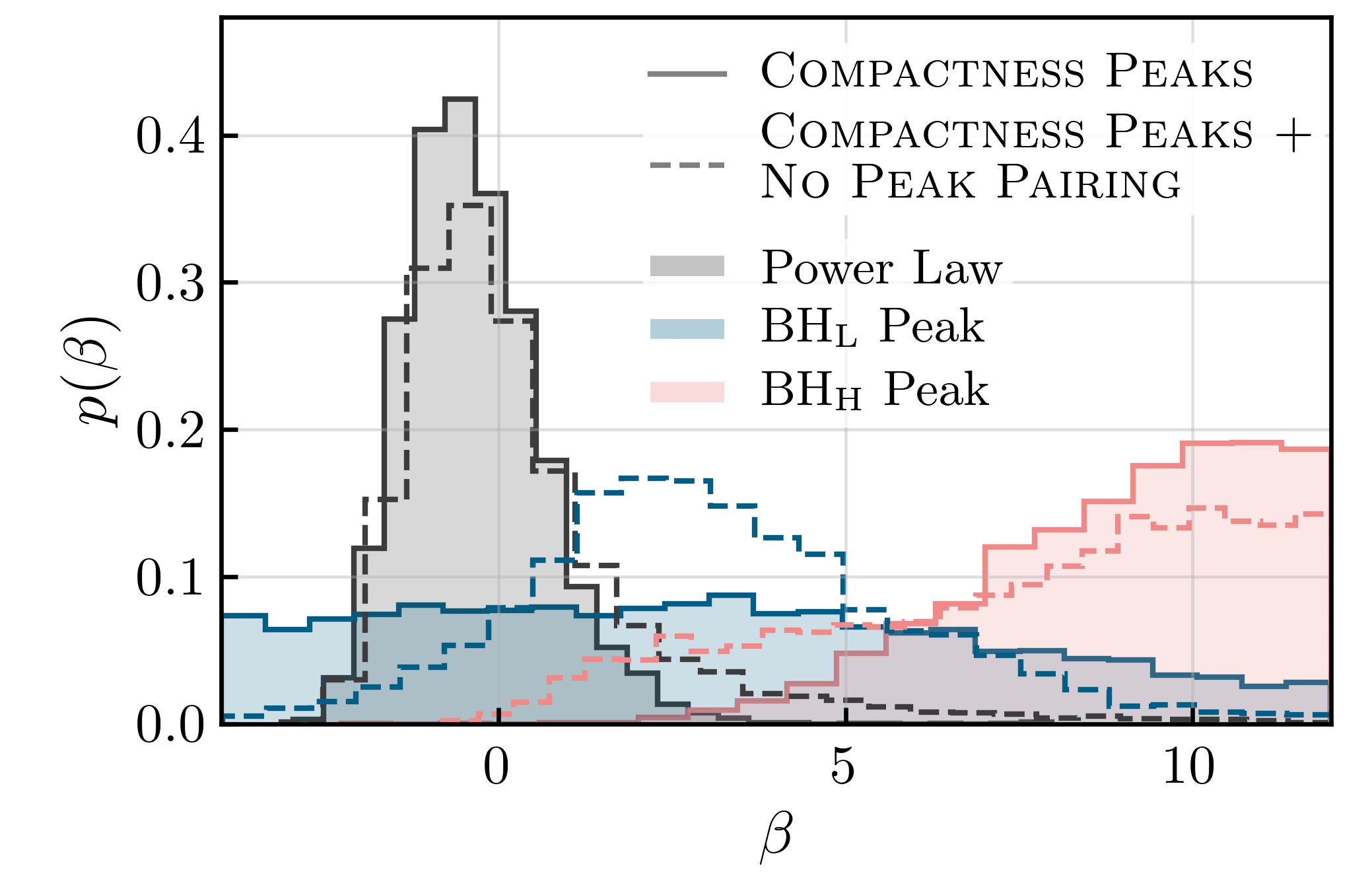}
    \caption{Slope index of the mass ratio distribution  $\mathrm{BH_L}$ Peak (blue), $\mathrm{BH_H}$ Peak (pink) and \textsc{Power Law} (black) components using the \cpeak (solid lines) and \cpeak \textsc{+ No Peak Pairing} (dashed lines) which has relaxed conditions on the pairing of $m_1$ and $m_2$.}
    \label{fig:betas}
\end{figure}

\begin{figure}
    \centering
    \includegraphics{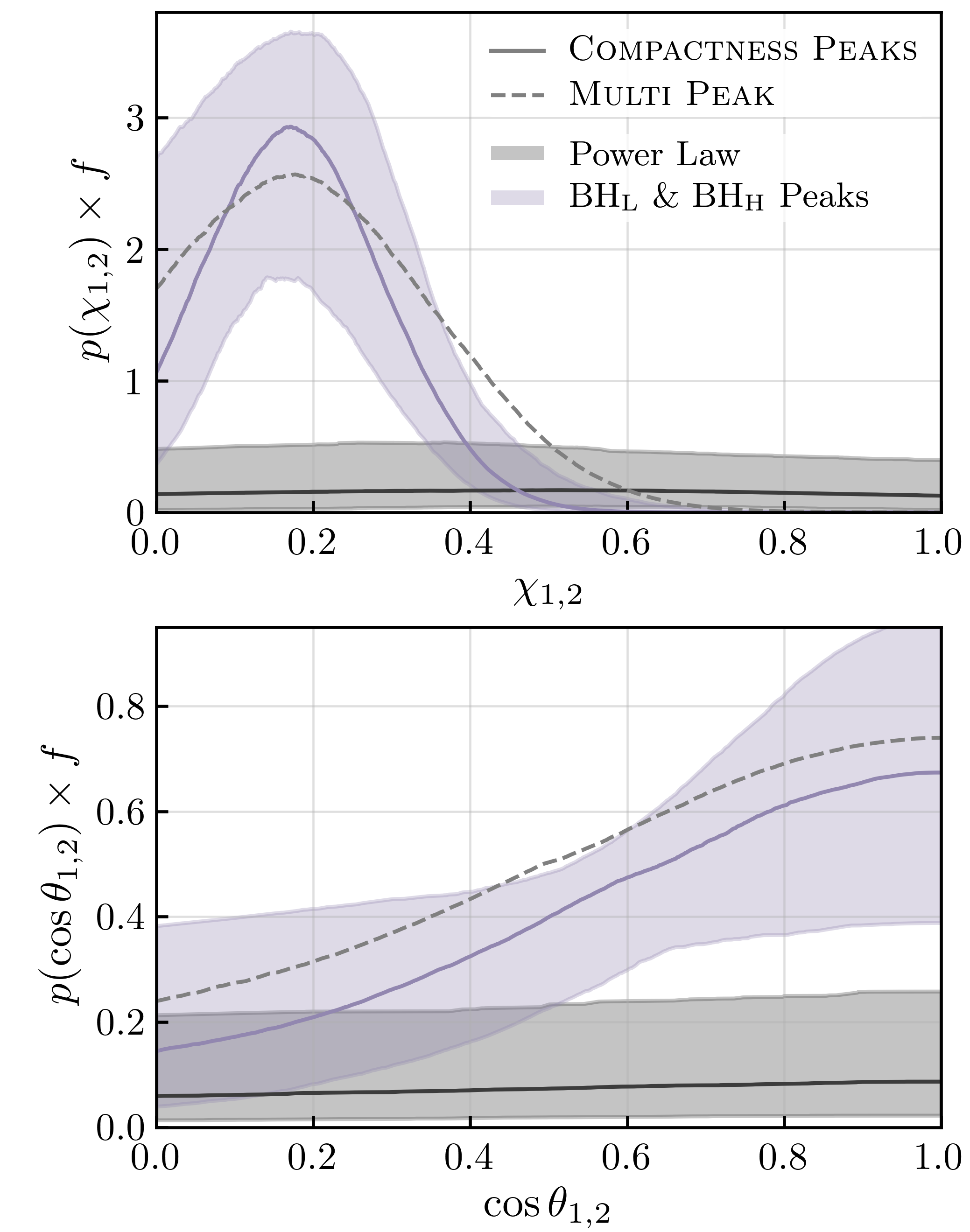}
    \caption{Population distributions for spin magnitude ($\chi$, top panel) and spin orientation ($\cos\theta$, bottom panel) using the \cpeak model for binary black holes mergers scaled by the fraction of events in the  $\mathrm{BH_L}$ and $\mathrm{BH_H}$ \textsc{Peaks} (purple), and \textsc{Power Law} (black) components.
    The solid curve is the median of the distribution and the shaded regions represent the 90\% credible interval.
    The dashed line represents the median of the distribution obtained from the \mpeak model analysis, which has the same spin distribution for all components.}
    \label{fig:spin_dist}
\end{figure}

We analyse the population of binary black hole mergers and provide population predictive distributions (PPDs). The PPD given model is our best guess for the distribution of some source parameter $\theta$, averaged over the posterior for population parameters $\Lambda$ (see Table~\ref{tab:compactnesspeaks_priors} for descriptions),
\begin{align}
    p_\Lambda(\theta | d) = \int d\Lambda \, 
    p(\Lambda | d) \,
    \pi(\theta | \Lambda) .
\end{align}
We plot and highlight the components of \cpeak and compare these results to the \mpeak model, the parametric model with the strongest preference used in \cite{GWTC3popKAGRA:2021duu}.
\footnote{We use a variation of \mpeak model without smoothing ($\delta_m=0$) and different priors on the parameters. For details on this model, see Section B.4 in \cite{GWTC2popLIGOScientific:2020kqk}, the priors used are provided in Table \ref{tab:multipeak_priors}. We use the same spin and redshift model as used for \cpeak, but the same spin distribution is used for all components.}
This model is similar to \cpeak in that it has a powerlaw and two peaks in the primary mass component, however, it does not have different mass ratio or spin distributions for the different primary mass components, nor does it allow for the peaks to truncate at locations different to the powerlaw components minimum and maximum values.
We calculate the Bayes factor ($\mathcal{B}$) to provide a comparison of the models and determine which model is preferred by the data (Table \ref{tab:bf}).
We find that \cpeak model is strongly preferred over the \mpeak model by $\log_{10}\mathcal{B} = 4.63$ (i.e. preferred a by factor of $\sim 40000$), supporting the hypothesis in \cite{Schneider:2023mxe}.

In Figure \ref{fig:mass_dist} we show the \textit{astrophysical} distributions for $m_1$ in the top panel and $q$ in the bottom panel. 
We find that allowing for different mass ratio distributions in each component and pairing of the peaked regions in $m_1$ and $m_2$, the powerlaw component of the primary mass distribution becomes less steep with a slope index of $\alpha = 1.82^{+1.01}_{-1.5}$ in comparison to the \mpeak model $\alpha = 2.82^{+0.93}_{-0.98}$ at 90\% credibility.
We find that the edges of the component mass gap, defined by the maximum of the lower mass peak and the minimum of the upper mass peak are $m_1 = 10.92^{+1.62}_{-0.56} \Msun$  and $m_1 = 13.09^{+3.02}_{-2.67} \Msun$ respectively at 90\% credibility (see the posterior distributions in Figure \ref{fig:mass_hyperparams}).

We also find shifts in the fraction of binary black holes in each component with events in the powerlaw component moving to the peaked components when comparing the \cpeak and \mpeak models as shown in Figure \ref{fig:fracs}.
The 90\% credible intervals for the fraction of binaries in each component are $f_{\mathrm{PL}} = 0.15^{+0.29}_{-0.11}$, $f_{\mathrm{BH_L}} = 0.73^{+0.13}_{-0.26}$ and $f_{\mathrm{BH_H}} = 0.1^{+0.08}_{-0.05}$ for \cpeak and $f_{\mathrm{PL}} = 0.35^{+0.44}_{-0.28}$, $f_{\mathrm{BH_L}} = 0.56^{+0.25}_{-0.42}$ and $f_{\mathrm{BH_H}} = 0.08^{+0.1}_{-0.04}$ for \mpeak.

In Figure \ref{fig:betas}, we consider the slope index, $\beta$, of the mass ratio distribution for each component.
We find that the powerlaw component has no preference for $q\sim1$ but rather a uniform distribution, whereas the $\mathrm{BH_H~Peak}$ component strongly prefers more positive slopes (i.e. $q\sim1$).
The $\mathrm{BH_L~Peak}$ component appears to return the prior for $\beta$.
This result is because the truncation in mass ratio distribution due to the strict pairing of the peaks in $m_1$ and $m_2$ is driving the shape of the mass ratio distribution.
However, if we relax this restriction by letting,
\begin{align}
    \pi(q | m_1, \Lambda)_\mathrm{BH_L} \propto 
    \begin{cases}
    q^{{\beta}_\mathrm{BH_L}} & m_\mathrm{min}^\mathrm{PL} < m_2 < m_1 \\
    0 & \text{otherwise},
    \end{cases}
\end{align}
and 
\begin{align}
    \pi(q | m_1, \Lambda)_\mathrm{BH_H} \propto 
    \begin{cases}
    q^{{\beta}_\mathrm{BH_H}} & m_\mathrm{min}^\mathrm{PL} < m_2 < m_1 \\
    0 & \text{otherwise},
    \end{cases}
\end{align}
such that $m_2$ is not confined to be within the component defined in $m_1$ space, we find the $\mathrm{BH_L~Peak}$ components also prefer positive slopes for the mass ratio distribution.
This analysis also prefers values of $q \sim 1$ for the peak components.
This model variation is henceforth referred to as \cpeak \textsc{+ No Peak Pairing}.
We also find this model with a relaxed condition on the mass ratio model is slightly more preferred over the \cpeak model by $\log_{10}\mathcal{B} = 0.3$.
While this preference suggests that the strict pairing of $m_1$ and $m_2$ in the peaked component regions (as illustrated in Equation \ref{eq:mass_ratio}) may not be necessary to describe the population at present, the preference of $q \sim 1$ in the peaked components still remains and can be accounted for by the positive values of $\beta$ as shown in Figure \ref{fig:betas}.

Figure \ref{fig:spin_dist} shows that the populations in the powerlaw and compactness peaks have different spin properties, where the peaked components have lower spins than the powerlaw component and are generally more aligned with respect to the orbital angular momentum.
However, the result for the powerlaw component is somewhat prior driven (see Figure \ref{fig:spin_hyperparams}) given the small fraction of binary black holes,  $f \sim 0.15$, in the powerlaw component.
We do observe that the powerlaw component has accounted for events with higher spins and very low spins, with reduced support for these values in the peaked distributions in comparison to the \mpeak model.

We also considered a few variations of the \cpeak and \cpeak \textsc{+ No Peak Pairing} without separate spin distributions (i.e. the same values for the spin parameters $\mu_\chi$, $\sigma_\chi$, $\sigma_t$, $\zeta$) for the powerlaw and peak populations (Table \ref{tab:bf}).
We found that support for these models is significantly driven by allowing for separate spin populations ($\log_{10}\mathcal{B} \sim 1-3$).
Interestingly, the degree of increase in model support with and without separate spin populations varies depending on the mass ratio distribution of the model.
This result suggests that mass and spin are not strictly independent parameters and a possible consequence of the $\chi_\mathrm{eff}-q$ correlation \citep{Callister:2021fpo, Adamcewicz:2022hce}.

\begin{table*}
\centering
\renewcommand{\arraystretch}{1.5}
\begin{tabular}{p{13cm} c c}
    \hline
    Model & $ \log_\mathrm{10}{\cal B}$ & $\Delta \log_\mathrm{10}{\cal L}_\mathrm{max}$ \\
    \hline\hline
\cpeak & $4.63$ & $6.84$ \\
\cpeak (w/ same $\mu_\chi$, $\sigma_\chi$, $\sigma_t$, $\zeta$) & $3.78$ & $6.02$ \\\hline
\cpeak \textsc{+ No Peak Pairing} & $4.93$ & $7.49$ \\
\cpeak \textsc{+ No Peak Pairing} (w/  same $\beta$ ) & $4.78$ & $6.22$ \\ 
\cpeak \textsc{+ No Peak Pairing} (w/ same $\mu_\chi$, $\sigma_\chi$, $\sigma_t$, $\zeta$)  & $2.88$ & $5.88$ \\ 
\cpeak \textsc{+ No Peak Pairing} (w/ same $\beta$, $\mu_\chi$, $\sigma_\chi$, $\sigma_t$, $\zeta$)  & $1.91$ & $1.51$ \\ \hline
\mpeak ($\delta_m = 0$) & ~~$0.0$ & ~~$0.0$ \\
\textsc{Powerlaw Peak} ($\delta_m = 0$) & $-1.97$ & $-3.44$ \\

    \hline
\end{tabular}
\caption{Log Bayes factors and maximum log likelihood differences for models compared to the \mpeak model \citep{GWTC2popLIGOScientific:2020kqk}.
We also include results for the \textsc{PowerLaw Peak} \citep{Talbot:2018cva}, a commonly used model for the population analysis of binary black holes \citep[e.g.][]{GWTC3popKAGRA:2021duu}
}
\label{tab:bf}
\end{table*}

\section{Discussion}\label{sec:discussion}

Motivated by the hypothesis presented in \cite{Schneider:2023mxe} for the presence of a gap in the component mass spectrum of merging binary black holes, we analysed the events in \cite{GWTC3popKAGRA:2021duu} using a population designed to capture the structure of the compactness peaks and the degree of pairing of the peak regions in $m_1$ and $m_2$.
We also explored whether there is a subpopulation of events that have properties different to the binary black holes we expect from binary stripped stars (e.g. formed via dynamical evolution).

We find that there is a preference for the lower mass peak to drop off sharply at $m_1 = 10.92^{+1.62}_{-0.56} \Msun$ and the upper mass peak to turn on at $m_1 = 13.09^{+3.02}_{-2.67} \Msun$ (90\% credible intervals). 
The location of the possible edges of the peaks are consistent with predictions from \cite{Schneider:2023mxe} with a possible gap, but, there is no \textit{clear} evidence for a gap-like feature given the posteriors of the edges of the peaks overlap.
If we consider the case where there is no powerlaw component to sweep up the ``pollutants'' (i.e. only the compactness peaks), support for a gap-like structure decreases, with the edges at $m_1=10.66^{+1.28}_{-0.44} \Msun$ and $m_1=12.1^{+1.78}_{-1.95} \Msun$ (90\% credible intervals).

We also find mild support for the peaked populations to have a different spin distribution to the population defined by the powerlaw component.
This result may hint at the possibility of the powerlaw component events being formed via other pathways to that of binary stripped stars in isolation, with $4-44\%$ of binaries in the powerlaw component compared to $56-96\%$ in the peaked components, with the majority of binaries, $48-87\%$, in the low mass peak (90\% credible intervals).
However, more events are needed to constrain these results.
With events from the fourth LVK observing run (O4), which includes KAGRA \citep{KAGRA:2018plz} observatory in the gravitational-wave network, we expect our sample size to grow by a factor of $\sim 3$ \citep{LVK:2018LRR....21....3A, Iacovelli:2022bbs, Kiendrebeogo:2023hzf}.
This increase may help constrain the parameters defining the powerlaw component and the edges of the compactness peaks.

Recent studies with a focus on looking for hierarchical mergers in the population find evidence that the spin properties of observed binary black holes changes at $m \approx 40-50 \Msun$ \citep[e.g.][]{Antonini:2024het, Pierra:2024fbl}.
In our work, we find that the upper mass compactness peak turns off at $m \approx 37 - 42 \Msun$, which appears to be consistent with these findings.
This result is not driven by the spin information for each component.
By fixing each mass component to have the same spin distribution, the turn off is at $m \approx 37 - 41 \Msun$.
However, there is a preference for the powerlaw component to have a different spin distribution to the compactness peak populations.
Given $\chi_\mathrm{eff}$ is a better constrained parameter, it would be interesting to see the variation of $\chi_\mathrm{eff}$ across these different populations and how it compares to our analysis with component spins.

In the late preparation stages of this paper, the authors became aware of a similar study exploring the proposed gap in the component masses \citep{Adamcewicz:2024jkr}.
In their work, they aim to answer the question: is there a lack of black holes between $m \approx 10-15 \Msun$?
To do this they directly model a dip-like feature in the component mass distribution using a notch filter, the same as proposed in \cite{Farah:2021qom}, to explore the gap between the masses of neutron stars and black holes.
Similar to our model, they also use two peaks that surround this gap.
They do not find evidence for a dip-like feature in the distribution.
Considering there are events ``polluting the gap'', and given the mass posteriors uncertainties are generally larger than the width of the gap, this result is not unexpected.
They also conclude that the dip-like feature will likely not reveal itself with an increased catalogue of events from O4.
In our work, we aimed to answer a slightly different question: is there a lack of black holes between $m \approx 10-15 \Msun$, \textit{given} the expectation of pollution in this region from binary black hole formed via other channels?
These two approaches lead to slightly differing conclusions, with support for the edges of the possible gap-like feature more pronounced in our work, and strong support for our \citeauthor{Schneider:2023mxe} inspired model ($\log_{10}\mathcal{B} = 4.63$).
However, both studies conclude that there is no \textit{clear} evidence for a gap-like feature in component mass.

In addition to this, the requirement of $q = m_2/m_1 \leq 1$ may also be impacting our results. 
This prior restriction results in the skewed posteriors we observe for some events where the mass ratio is close to unity. 
This effect is particularly evident in the events with masses $< 15 \Msun$, where slight shifts in mass lead to much more unequal mass ratios.
To mitigate this, future analyses extending this prior range for the individual events may help resolve the structure and edges of the compactness peaks.
Extending this, it may be interesting to study this same population with a spin sorting approach rather than a mass sorting approach, where the objects are ordered by the spin magnitude rather than the mass. 
This approach can also constrain spin information and has the most impact for events of $q \sim 1$ \citep[e.g.][]{Biscoveanu:2020are}

We highlight that in recent years, there has been a move towards data-driven models \citep[e.g.][]{Edelman:2021zkw, Tiwari:2023xff, Callister:2023tgi} for the analysis of compact binary populations.
These data-driven methods are valuable for finding structure in distribution without strong assumptions of the shape of the underlying distribution.
However, astrophysically-informed parametric models provide a clear method for astrophysical interpretations of the features observed and are ideal for model comparison.
In this data-rich era, both data-driven and phenomenological models are needed: one to identify the possible new structure and one to interpret the astrophysical implications for the features we observe.\\

\section{Acknowledgements}

Galaudage and Lamberts are supported by the ANR COSMERGE project, grant ANR-20-CE31-001 of the French Agence Nationale de la Recherche. The authors are grateful for computational resources provided by the LIGO Laboratory and supported by National Science Foundation Grants PHY-0757058 and PHY-0823459. This material is based upon work supported by NSF's LIGO Laboratory which is a major facility fully funded by the National Science Foundation

\begin{appendix}

\section{Population model details}

In this section we provide details about the population models described above in Section~\ref{subsec:methods_population}.
We include a summary of the parameters for the \cpeak and \mpeak models and the prior distribution used for each parameter in Tables \ref{tab:compactnesspeaks_priors} and \ref{tab:multipeak_priors} respectively.
The prior distributions are indicated using abbreviations: for example, U$(0,1)$ translates to uniform on the interval $(0,1)$.
The full \cpeak model is given by,

\begin{align}
    \pi & ( m_1, q, \chi_1, \chi_2, \cos\theta_1, \cos\theta_2 |\Lambda) \nonumber \\ &
    = (1-f_\mathrm{peaks}) \times \pi(m_1 |\Lambda)_\mathrm{PL} \times  \pi(q | m_1, \Lambda)_\mathrm{PL}  \nonumber \\ & \times \pi(\chi_{1,2} |\Lambda)_\mathrm{PL} \times \pi(\cos\theta_{1,2} |\Lambda)_\mathrm{PL} ~ \nonumber \\ &
    + \pi(\chi_{1,2} |\Lambda)_\mathrm{peaks} \times \pi(\cos\theta_{1,2}|\Lambda)_\mathrm{peaks}  \nonumber \\ & \times [ f_\mathrm{peaks} f_1 \times p(m_1 |\Lambda)_\mathrm{BH_L} \times  \pi(q | m_1, \Lambda)_\mathrm{BH_L} \times \nonumber \\ & +
    f_\mathrm{peaks} (1-f_1) \times p(m_1 |\Lambda)_\mathrm{BH_H} \times  \pi(q | m_1, \Lambda)_\mathrm{BH_H}
    ].
\end{align}

\begin{table*}
    \centering
    \renewcommand{\arraystretch}{1.5}
    \begin{tabular}{ c p{11.9cm} p{1mm} p{1.6cm} }
        \hline
        {\bf Parameter} & \textbf{Description} &  & \textbf{Prior} \\\hline\hline
        $f_\mathrm{peaks}$ & Fraction of black holes in the Gaussian components of the $m_1$ distribution. &  & U(0, 1) \\ 
        $f_1$ & Fraction of black holes in the lower-mass Gaussian component of the Gaussian components of the $m_1$ distribution. &  & U(0, 1) \\ \hline
        $\alpha$ & Spectral index for the power-law of the $m_1$  distribution. &  & U(-4, 8) \\
        $m_\mathrm{min}^\mathrm{PL}$ [$\Msun$] & Minimum mass of the power-law component of the $m_1$ distribution. &  & U(2, 10)\\
        $m_\mathrm{max}^\mathrm{PL}$ [$\Msun$] &  Maximum mass of the power-law component of the $m_1$ distribution. &  & U(60, 100)\\
        $\beta_\mathrm{PL}$ & Spectral index for the power-law of the $q$ distribution paired with the power law component of the $m_1$ distribution. &  & U(-4, 12) \\ \hline
        $\mu_\mathrm{BH_L}$ [$\Msun$] & Mean of the lower-mass Gaussian component in the $m_1$ distribution.  &  & U(8, 12) \\
        $\sigma_\mathrm{BH_L}$ [$\Msun$] & Width of the lower-mass Gaussian component  in the $m_1$ distribution.  &  & U(1,4)\\
        $m_\mathrm{min}^{\mathrm{BH_L}}$ [$\Msun$]& Minimum mass of the lower-mass Gaussian component of the $m_1$ distribution. &  & U(5, 8)\\
        $m_\mathrm{max}^{\mathrm{BH_L}}$ [$\Msun$]&  Maximum mass of the lower-mass Gaussian component of the $m_1$ distribution. &  & U(10, 15)\\
        $\beta_\mathrm{BH_L}$ & Spectral index for the power-law of the $q$ distribution paired with the power law component of the $m_1$ distribution. &  & U(-4, 12) \\
        \hline
        $\mu_\mathrm{BH_H}$ [$\Msun$] & Mean of the upper-mass Gaussian component in the $m_1$ distribution.  &  & U(12, 50) \\
        $\sigma_\mathrm{BH_H}$ [$\Msun$] & Width of the upper-mass Gaussian component in the $m_1$ distribution. &  & U(4, 20) \\
        $m_\mathrm{min}^{\mathrm{BH_H}}$ [$\Msun$]& Minimum mass of the upper-mass Gaussian component of the $m_1$ distribution. &  & U(10, 15)\\
        $m_\mathrm{max}^{\mathrm{BH_H}}$ [$\Msun$]&  Maximum mass of the upper-mass Gaussian component of the $m_1$ distribution. &  & U(30, 50)\\
        $\beta_\mathrm{BH_H}$ & Spectral index for the power-law of the $q$ distribution paired with the upper-mass Gaussian component of the $m_1$ distribution. &  & U(-4, 12) \\
        \hline
        $\mu_\chi^X$ & Mean of the spin magnitude distribution for a given component.  &  & U(0, 1) \\
        $\sigma_\chi^X$ &  Width of the spin magnitude distribution for a given component &  & U(0.1, 2) \\
        $\sigma_t^X$  & Width of the spin orientation distribution for the subpopulation aligned with respect to the orbital angular momentum for a given component  &  & U(0.1, 4) \\
        $\zeta^X$  & Fraction of binary black holes in the subpopulation aligned with respect to the orbital angular momentum for a given component &  & U(0, 1) \\
        \hline
    \end{tabular}
    \caption{
    Summary of parameters in \cpeak population model. The prior ranges are motivated by the results in \cite{Schneider:2023mxe}
    }
  \label{tab:compactnesspeaks_priors}
\end{table*}

\begin{table*}
    \centering
    \renewcommand{\arraystretch}{1.5}
    \begin{tabular}{ c p{11.9cm} p{1mm} p{1.6cm} }
        \hline
        {\bf Parameter} & \textbf{Description} &  & \textbf{Prior} \\\hline\hline
        $\alpha$ & Spectral index for the power-law of the $m_1$  distribution. &  & U(-4, 12) \\
        $m_\mathrm{min}$ [$\Msun$] & Minimum mass of the power-law component of the $m_1$ distribution. &  & U(2, 10)\\
        $m_\mathrm{max}$ [$\Msun$] &  Maximum mass of the power-law component of the $m_1$ distribution. &  & U(30, 100)\\
        $\beta$ & Spectral index for the power-law of the $q$ distribution paired with the power law component of the $m_1$ distribution. &  & U(-4, 12) \\ \hline
        $f_\mathrm{peaks}$ (or $\lambda$) & Fraction of black holes in the Gaussian components of the $m_1$ distribution. &  & U(0, 1) \\ 
        $f_1$ (or $\lambda_1$) & Fraction of black holes in the lower-mass Gaussian component of the Gaussian components of the $m_1$ distribution. &  & U(0, 1) \\ \hline
        $\mu_{m,1}$ [$\Msun$] & Mean of the lower-mass Gaussian component in the $m_1$ distribution.  &  & U(5, 20) \\
        $\sigma_{m,1}$ [$\Msun$] & Width of the lower-mass Gaussian component in the $m_1$ distribution.  &  & U(1,5)\\
        \hline
        $\mu_{m,2}$ [$\Msun$] & Mean of the upper-mass Gaussian component in the $m_1$ distribution.  &  & U(20, 50) \\
        $\sigma_{m,2}$ [$\Msun$] & Width of the upper-mass Gaussian component in the $m_1$ distribution. &  & U(1, 10) \\
        \hline
    \end{tabular}
    \caption{
    Summary of parameters in \mpeak model. This model, with similar prior ranges, was the most preferred parametric model used in \cite{GWTC3popKAGRA:2021duu}; see Appendix D.
    }
  \label{tab:multipeak_priors}
\end{table*}

\section{Hyperparameter posteriors}

\begin{figure*}
    \includegraphics[width=0.99\columnwidth]{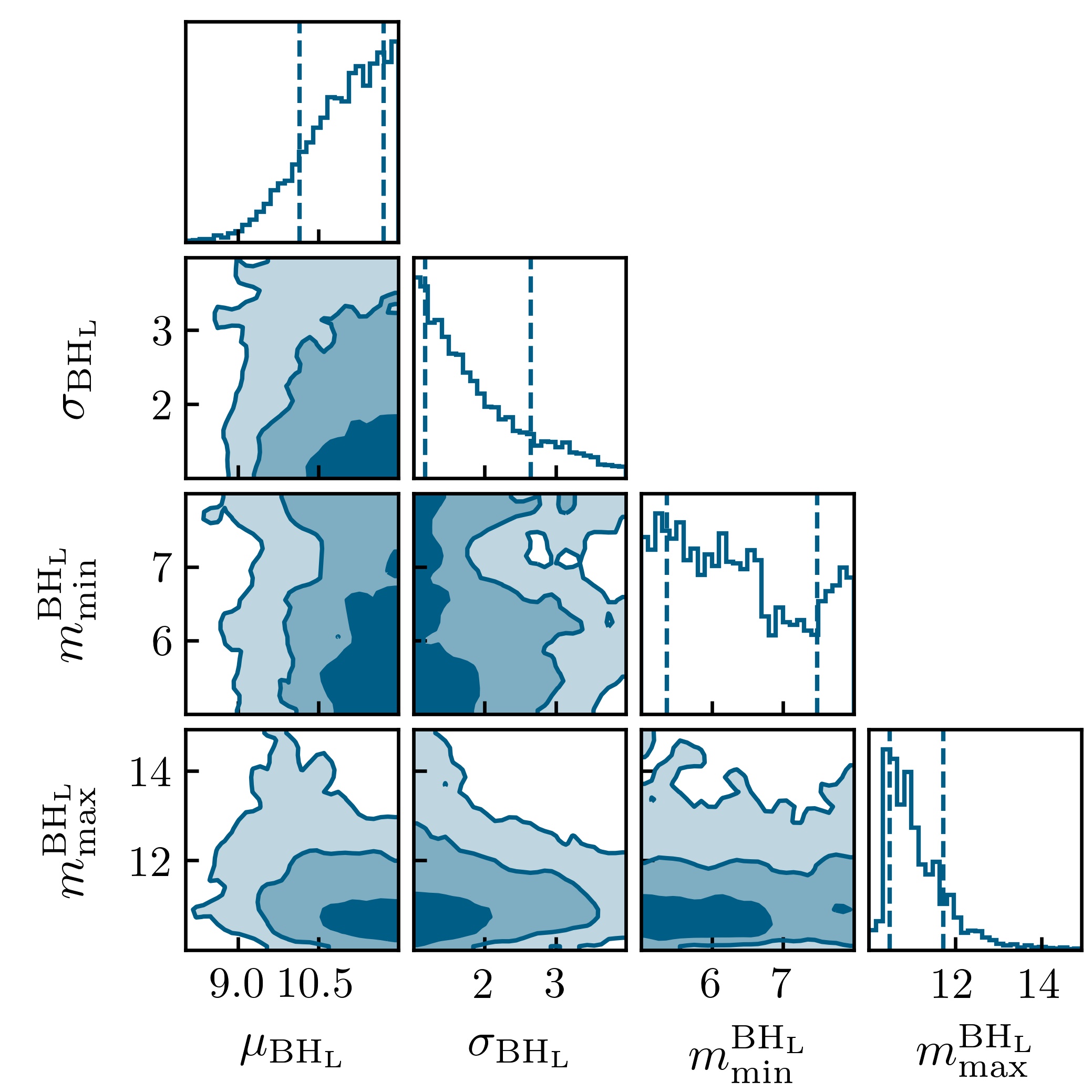}
    \includegraphics[width=0.99\columnwidth]{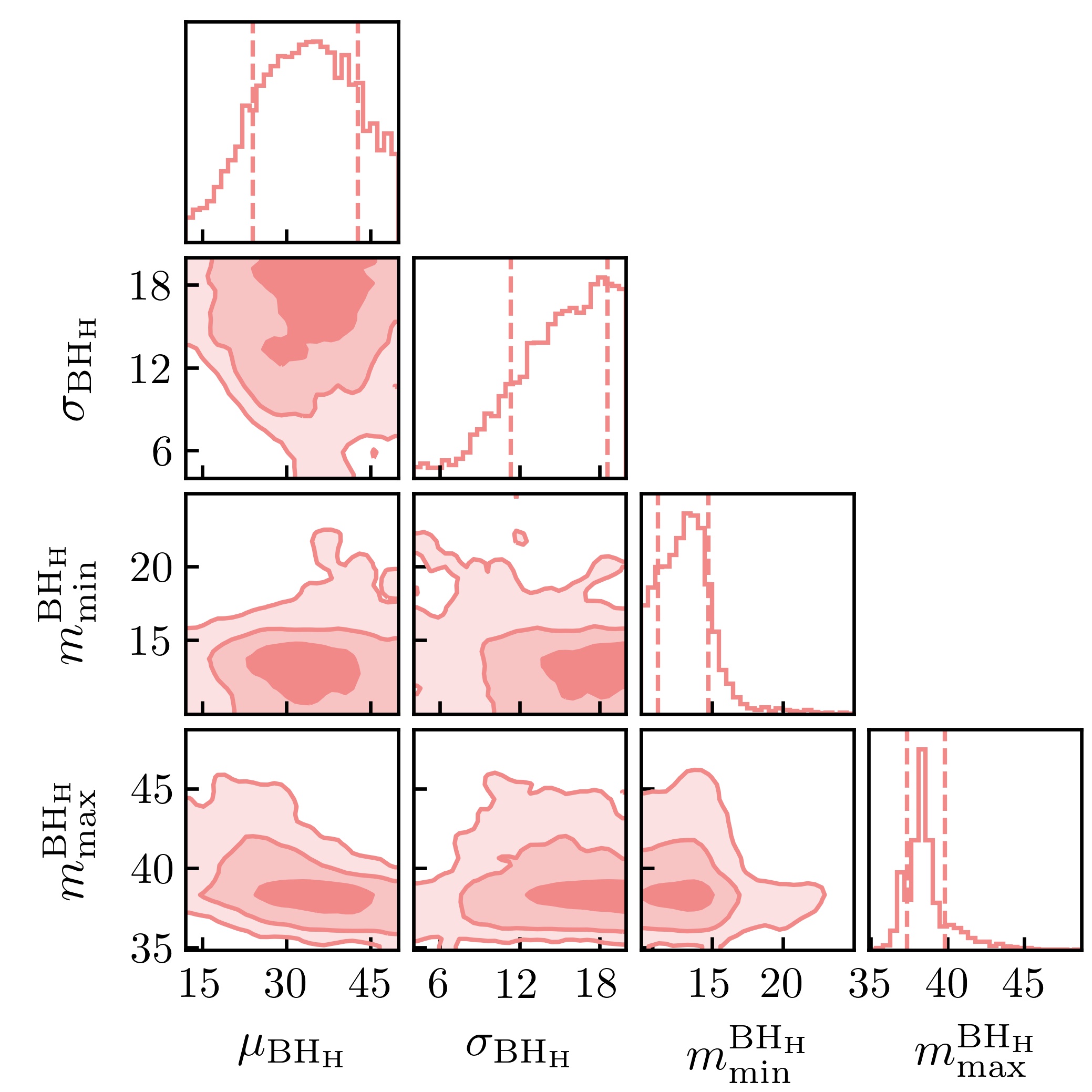}
    \caption{Mass hyperparameter posteriors of the peak components in the \cpeak model. \textit{Left panel}: the lower mass ($\mathrm{BH_L}$) peak where $\mu_\mathrm{BH_L}$ is the mean of the peak, $\sigma_\mathrm{BH_L}$ is the width of the peak and $m_\mathrm{min}^{\mathrm{BH_L}}$ and $m_\mathrm{max}^{\mathrm{BH_L}}$ are the minimum and maximum mass of the peak. \textit{Right panel}: the upper mass ($\mathrm{BH_L}$) peak where $\mu_\mathrm{BH_H}$ is the mean of the peak, $\sigma_\mathrm{BH_H}$ is the width of the peak and $m_\mathrm{min}^{\mathrm{BH_H}}$ and $m_\mathrm{max}^{\mathrm{BH_H}}$ are the minimum and maximum mass of the peak. 
    The credible intervals on the 2D posterior distributions are at $1 \sigma$, $2 \sigma$ and $3 \sigma$, using increasingly light shading.
    The intervals on the 1D posterior distributions are at $1 \sigma$.
    }
    \label{fig:mass_hyperparams}
\end{figure*}

\begin{figure*}
    \includegraphics[width=0.99\columnwidth]{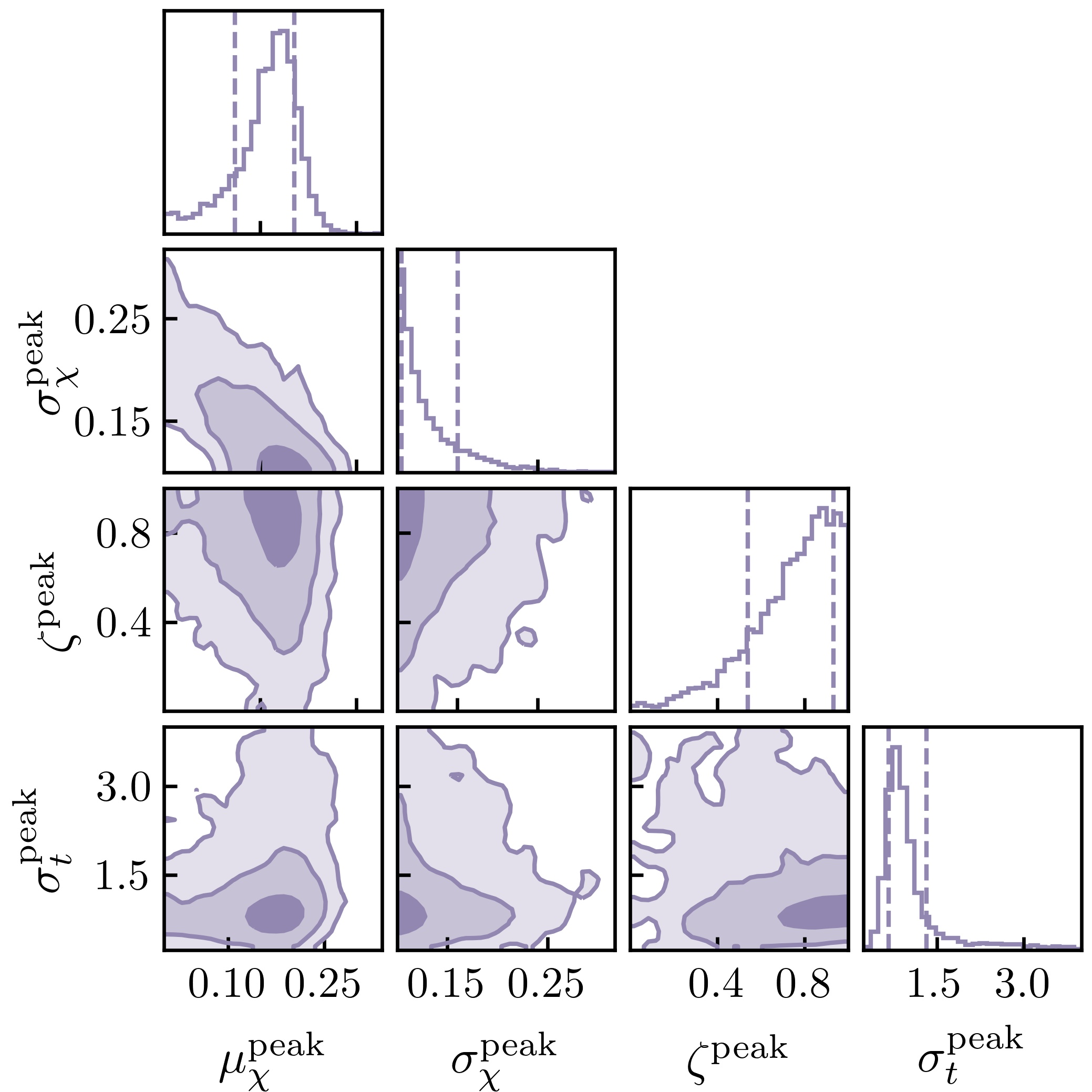}
    \includegraphics{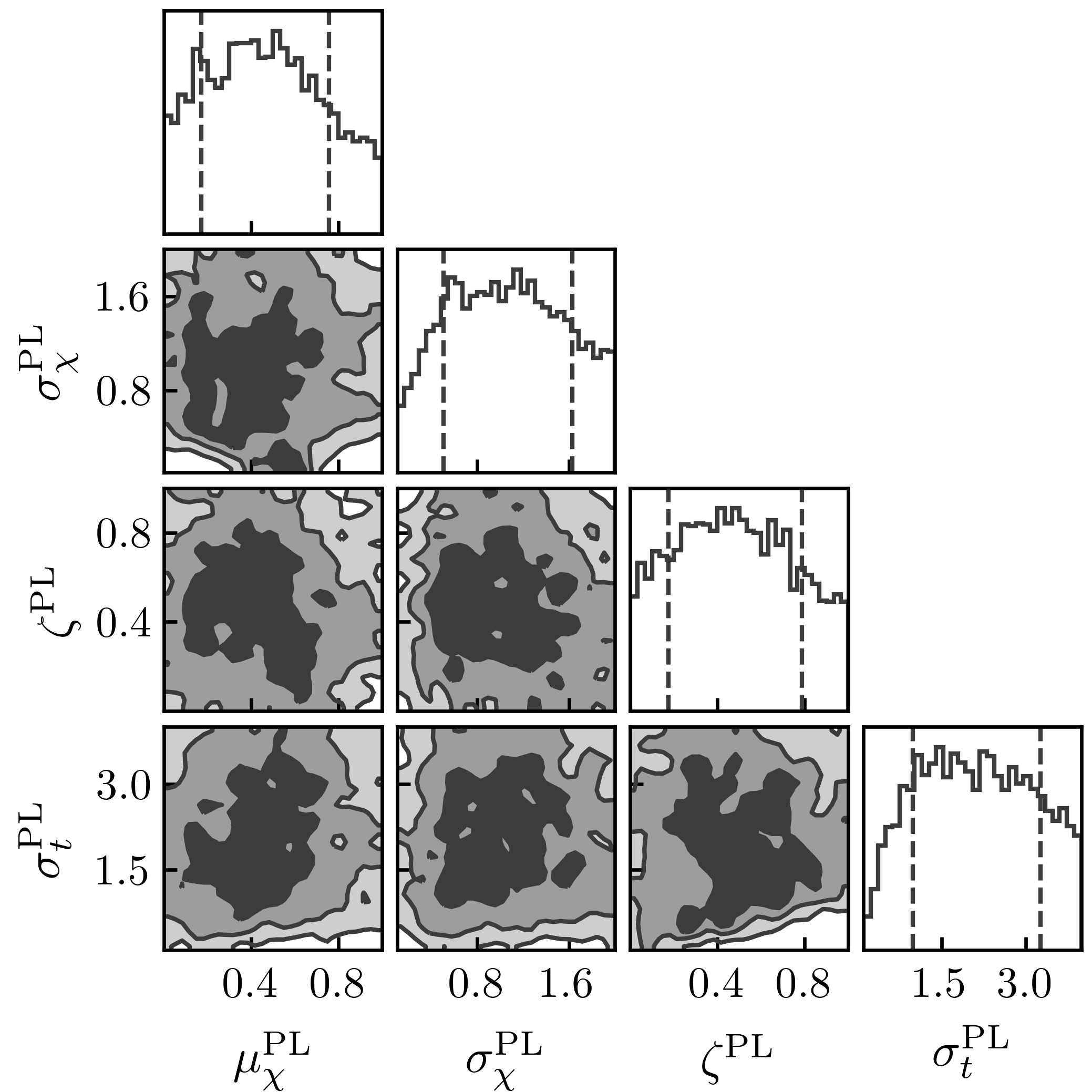}
    \caption{Spin hyperparameter posteriors of the \cpeak model, where $\mu_\chi$ is the mean of the spin magnitude distribution, $\sigma_\chi$ is the width of the spin magnitude distribution, $\zeta$ is the fraction of binary black holes in the subpopulation aligned with respect to the orbital angular momentum and $\sigma_t$ is the width of the spin orientation distribution for the subpopulation aligned with respect to the orbital angular momentum for a given component. \textit{Left panel}: The spin parameters paired with the $\mathrm{BH_L}$ and $\mathrm{BH_H}$ peaks \textit{Right panel}: The spin parameters paired with the powerlaw ($\mathrm{PL}$) component. 
    The credible intervals on the 2D posterior distributions are at $1 \sigma$, $2 \sigma$ and $3 \sigma$, using increasingly light shading.
    The intervals on the 1D posterior distributions are at $1 \sigma$.}
    \label{fig:spin_hyperparams}
\end{figure*}

In this section we provide the hyperparameter posteriors for the parameters governing the peak components of the mass distribution (Figure \ref{fig:mass_hyperparams}) and the spin parameters for the peaked and powerlaw components (Figure \ref{fig:spin_hyperparams}).

\end{appendix}

\bibliographystyle{aa}
\bibliography{main}

\end{document}